\documentclass[12pt,oneside,a4paper]{paper}
\usepackage{graphicx, natbib}

\newcommand{\ms}{\mbox{m s$^{-1}~$}}

\begin{document}

\title{Comment on "Stellar activity 
masquerading as planets in the habitable
zone of the M dwarf Gliese 581"
\footnote{Pre-print version - Published as a Science 'Technical comment' on 6 March 2015: 
Full reference : Science 6 March 2015. Vol. 347 no. 6226 p. 1080, DOI:
10.1126/science.1260796.
Correspondence to: guillem.anglada@gmail.com}
}
\author{
Guillem Anglada-Escud\'e $^{1,2}$
Mikko Tuomi $^{2}$\\
\normalsize $^{1}$ School of Physics and Astronomy,
Queen Mary University of London\\
\normalsize
327 Mile End Rd. London, United Kingdom\\
\\
\normalsize
$^{2}$
Centre for Astrophysics Research,
University of Hertfordshire\\
\normalsize
College Lane, AL10 9AB, Hatfield, UK\\
}
\bibliographystyle{mn2e}



\maketitle

\label{firstpage}

\begin{abstract}

Robertson et al.(Reports, July 25 2014, p440-444)(1) claimed that
activity-induced variability is responsible for the Doppler signal of the
proposed planet candidate GJ 581d. We point out that their analysis using
periodograms of residual data is incorrect, further promoting inadequate tools.
Since the claim challenges the viability of the method to detect exo-Earths, we
urge for a correct re-analysis (provided as an appendix in pre-print version).

\end{abstract}


GJ~581d was the first planet candidate of a few Earth masses reported in the
circumstellar habitable zone of another star (1). It was detected by measuring
the radial velocity variability of its host star using High Accuracy Radial
Velocity Planet Searcher (HARPS) (1,2). Doppler time series are usually modeled
as the sum of Keplerian signals plus additional effects (e.g., correlations with
activity). Detecting a planet candidate consists of quantifying the improvement
of a merit statistic when one signal is added to the model. Approximate methods
are often used to speed up the analyses, such as computing periodograms on
residual data. Even when models are linear, correlations exist between
parameters. Similarly, statistics based on residual analyses are biased
quantities and cannot be used for model comparison. 

\begin{figure*}
\center
\includegraphics[angle=0, width=0.85\textwidth, clip]{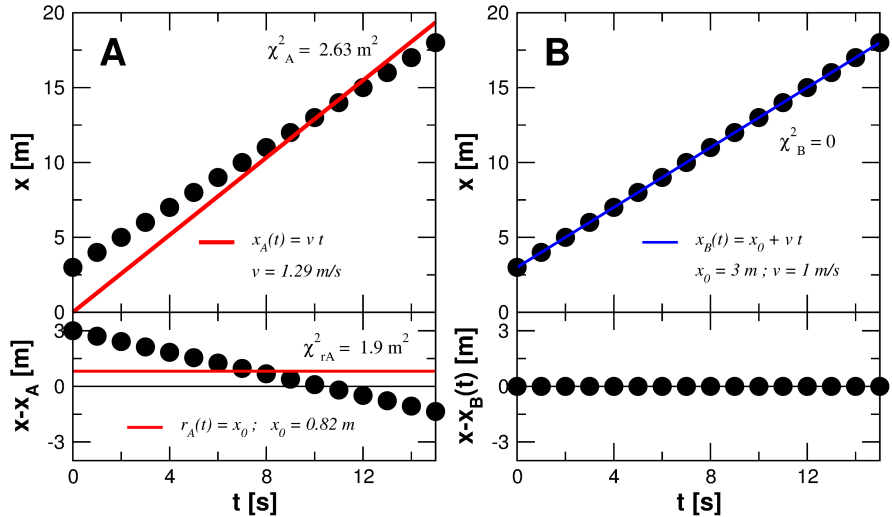}
\caption{This example illustrates why residual statistics must not be used to  assess
significances in multi-parametric fits to data. We want to know  whether a
constant $x_0$ is needed to model the position $x$ of a body as  a function of
time $t$ (black dots). Top left panel represents the fit to  the null hypothesis
(model A) which only includes a velocity term (red line).  Bottom left shows the
residuals $r_A = x-x_A(t)$ to model A. An attempt to fit a  model with a
constant offset to $r_A$ is shown as a red line. Top right panel  represents a fit
adjusting all the parameters simultaneously (model B, top right panel, blue
line), producing the largest possible reduction in $\chi^2$.} 
\label{fig:graph}
\end{figure*}

A golden rule in data-analysis is that the data should not be corrected, but it
is our model which needs improvement. The inadequacy of residual analyzes can be
illustrated using a trivial example (Figure 1). Let's assume 16 measurements of
the position of an object ($x$) as a function of time($t$) and no uncertainties.
We are interested in its velocity but we need to decide whether constant offset
$x_0$ is needed to model the motion. Model A (null hypothesis) consists consists
in $x_A(t) = v t$ , where $v$ is the only free parameter, and the alternative
Model B is $x_B(t)=x_0 + vt$ . The question is whether including x$_0$ is
justified given the improvement of a statistic that we define a 
$\chi^2=\frac{1}{N}\sum_{i=1}^{N} \left(x_i-x(t_i)\right)^2$. Left panel in
Figure 1 illustrates an inappropriate procedure which consist on adjusting Model
A, and then deciding whether a constant $x_0$ is needed to explain the residuals
(bottom left panel). Since such residuals are far from a constant shift, the
reduction of $\chi^2$ is not maximal and the fit to a constant offset
unsatisfactory. By subtracting model A from the data, we have created a new
time-series which is no longer representative of the original one. A more
meaningful procedure consists in comparing model A to a the global fit to all
the parameters of model~B (top right panel) achieving maximal improvement of our
statistic.

Similarly, the analysis in Robertson et al.(3) only shows that the signal of
GJ~581d is not present in their new residual time-series. Their procedure is
summarized as follows. Their figures 1 and S3 were used to suggest RV/H$_\alpha$
correlations contaminating the measurements. After subtracting those
correlations and the first three planets, periodograms(4-6) were applied to the
residuals to show that GJ 581d fell below the detectability threshold. While the
signal of GJ 581d is $K\sim 1.6$ m/s, the apparent variability induced by the
RV/H$_\alpha$ correlations is 5 m/s peak-to-peak, and the scatter around the
fits is at the $1.5-2$ m/s level. Subtracting those correlations biased the
residuals by removing a model that likely included contributions from real
signals and additional noise was added due to the scatter in the RV/H$_\alpha$
relations. All things considered, the disappearance of GJ~581d in such residual
data is not surprising. Following Fig.~1, a simultaneous fit of the 30+
parameters involved would be needed to reach meaningful conclusions. Although
there may be substantial RV/H$_\alpha$ correlations, a global optimization
analysis may not support that GJ 581d is better explained by activity. A
complete analysis will be presented elsewhere \textit{(see Appendix on this
preprint version).}

We argue that the results of Robertson et al.(3) come from the improper use of
periodograms on residual data as they implement the same flawed procedure
illustrated in Figure 1. Despite such periodograms are useful to provide
quick-look analyses, their inadequacy to the task has been abundantly discussed
in the literature(7--12). Explicitly, derived false alarm probabilities would be
representative only if a model with one-sinusoid and one offset is a sufficient
description of the data, measurements are uncorrelated, noise is normally
distributed, and uncertainties are fully characterized (5). Every single of
these hypothesis breaks down when dealing with Doppler residuals: the number of
signals in not known a priori, fits to data correlate residuals, and formal
uncertainties are never realistic. Proposed alternatives such as Monte Carlo
bootstrapping of periodograms(5) do not help either, as those methods ignore
correlations as well. Resulting biases can lead to significance assessments off
by several orders of magnitude. These issues were irrelevant when Doppler
amplitudes abundantly exceeded uncertainties. For example, an amplitude larger
than three times the uncertainties and more than 20 measurements easily leads to
false alarm probabilities smaller than 10$^{-6}$, which is much smaller than
usual thresholds at 1\%-0.1\%. For this reason, large biases were not
problematic in the early detection of gas giants (K$\sim 50$ m/s and 
$\sigma\sim 5$ m/s)(13), and it is the main reason why periodograms of residual
data are still wide-spread tools in Doppler analyses despite their inadequacy to
the task.

In summary, analysis of significance using residual data statistics leads to
incorrect significance assessments. While this has been a common practice in the
past, the problem is now exacerbated with signals closer to the noise and
increased model complexity. The properties of the noise can be included to the
model, but never subtracted from the data. This discussion directly impacts the
viability of the Doppler method to find Earth-like planets. While Earth causes a
0.1 m/s wobble around the Sun, the long-term stability of the most quiet stars
seems not better than 0.8 m/s(3). That is, activity induced variability can be
5-10 times larger than the signal. While global optimization does not provide
absolute guarantees of success, analyzes based on residual statistics are
certainly bound to failure. If activity poses an ultimate barrier to the
detection of small planets, strategic long-term plans concerning large projects
will need serious revision(14). It is of capital importance that analysis and
verification of multi-planet claims are properly done using global-optimization
techniques and by acquiring additional observations.
\\
\noindent
\textbf{Acknowledgments.} This work has been mostly supported by The Leverhulme
Trust through grant RPG 2014-281 - PAN-Disciplinary algORithms for data
Analysis. We thank H. R. A. Jones and R. P. Nelson for useful discussions and
support.
\\
\\
\noindent
\textbf{References and Notes}
\begin{enumerate}

\item
M. Mayor,X. Bonfils, T. Forveille, et al. The HARPS search for southern 
extra-solar planets. XVIII. An Earth-mass planet in the GJ 581 
planetary system. A\&A 507, 487A (2009)

\item
F. Pepe, C. Lovis, D. Segransan, et al. The HARPS search for Earth-like 
planets in the habitable zone. I. Very low-mass planets around HD 20794,
 HD 85512 and HD 192310. A\&A 534, A58 (2011)

\item
P. Robertson, S. Mahadevan, M. Endl, A. Roy. Stellar activity 
masquerading as planets in the habitable zone of the M dwarf 
Gliese 581. Science 345 (6195): 440-444 (2014)

\item
N. R. Lomb. Least-squares frequency analysis of unequally spaced 
data, Ap\&SS 39, 447L  (1976)

\item
A. Cummings. Detectability of extrasolar planets in radial velocity 
surveys, MNRAS 354, 1165C (2004)

\item
M. Zechmeister, M. Kuester,  The generalised Lomb-Scargle periodogram. 
A new formalism for the floating-mean and Keplerian periodograms A\&A 496, 577Z (2009)

\item
P.C. Gregory. A Bayesian Analysis of Extrasolar Planet Data for 
HD~73526, ApJ 631, 1198G (2005)

\item
R. P. Baluev. Assessing the statistical significance of 
periodogram peaks, MNRAS 385, 1279B, (2008)

\item
M. Tuomi. Bayesian re-analysis of the radial velocities of 
Gliese~581. Evidence in favour of only four planetary companions,  
A\&A 528L, 5T (2011)

\item
G. Anglada-Escude, M. Tuomi. A planetary system with gas giants 
and super-Earths around the nearby M dwarf GJ 676A. Optimizing 
data analysis techniques for the detection of multi-planetary 
systems, A\&A 548A, 58A (2012)

\item
R.D. Haywood,, A. Collier Cameron. D. Queloz, et al. Planets and 
stellar activity: hide and seek in the CoRoT-7 system, MNRAS 443, 
2517H (2014)

\item
R. Baluev. Accounting for velocity jitter in planet search 
surveys. MNRAS 393, 969B (2009)

\item
M. Mayor, D. Queloz. A Jupiter-mass companion to a solar-type 
star, Nature 378, 355M (1995)

\item
Several instruments aimed at precision better than 1 m/s are 
being proposed and/or in construction (e.g., ESPRESSO-VLT at 
the European Southern Observatory, TPF at the Hobby-Eberly 
Telescope, CARMENES at the Calar Alto Observatory, and SPIRou 
at the Canadian France Hawaii Telescope), because they are
 considered essential to detect and Earth-like planets or 
 confirm/characterize those detected by next planet-hunting 
 space missions (K2/NASA, TESS/NASA and PLATO/ESA). 

\end{enumerate}

\newpage

\begin{center}
\,
{\Huge Appendix}\\
\vskip 1cm
{\large (pre-print version only)}
\end{center}

\newpage
\appendix

\section{GJ~581d is not better explained by stellar activity}

We present a re-analysis of the data presented in \citet{robertson:2014}
(R14 hereafter) and show that the main conclusion of that manuscript ('GJ 581
does not exist', abstract quoting) is not supported by a gobal fit to the data.
The analysis is done by adding parameterized effects to the model (never
subtracting them) and using the maximum likelihood statistic as a
figure of merit (a generalization of the $\chi^2$ statistic). We limit ourselves
to a \textit{frequentist} framework, which is sufficient to illustrate the
perils of analise based on residual statistics. In all that follows, we use
exactly the same data as provided in R14 to show that the discrepant result
comes from basic statistical assumptions, and it is not a matter about the
quality or properties of the data.

In Section \ref{sec:model}, we outline the Doppler model used and show how the
correlations with activity indices are introduced in it. Section
\ref{sec:periodograms} reviews how to produce periodograms that account for the
presence of several free parameters in addition to the new signal of interest.
Results of the analysis are given in \ref{sec:results}. In Section
\ref{sec:timevariability} we argue that, although the correlations with the
I$_{H_\alpha}$ index are substantial, there is no clear evidence for
time-variability, and highlight the perils of applying arbitrary slicing to
datasets and fitting unconstrained correlation laws. Concluding remarks are
given in Section \ref{sec:conclusions}.

\subsection{Doppler and statistical model}\label{sec:model}

Our model to predict the radial measured radial velocity $v$ of a star
given the presence of k-planets is given by
\begin{eqnarray}
v\left[\vec{\theta}; t_i\right] = 
\gamma_I + \sum^k_p m[\vec{\theta}_p, t_i] + \dot{v}(t_i-t_0) + C_I \alpha_i 
\end{eqnarray}
\noindent where $t_i$ is the instant of each observation, $\gamma_I$ is a
constant offset of each instrument $I$ (or dataset), $\dot{v}$ is a term to
account for a long term secular acceleration common to all datasets, and the
usual Keplerian parameter of each planet candidate $p$ are consolidated in
$\vec{\theta_p}$ : Period $P_p$ in days, amplitude $K_p$ in \ms, eccentricity
$e_p$, argument of periastron $\omega_w$ in degrees, and initial mean $M_{0,p}$
in degrees. As discussed in R14 and already proposed in the past
\citep[eg.][]{queloz:2001, bonfils:2007}, some activity indices can linearly
correlate with spurious Doppler offsets. The rightmost term accounts for such
correlation and $\alpha_i$ is some simultaneaous activity measurement (eg.
$I_{H_\alpha}$ provided in R14 in this case). $C_I$ can
differ between instruments (correlations can be wavelength and resolution
dependent), so one $C_I$ is needed for each dataset $I$.  Since the mean value
of the activity index is not known, a linear correlation model should also
include a constant offset (eg. $v = C_I \alpha + b_I$, where $b_I$ should be a
free parameter as well). However note that such constant is automatically
absorbed by $\gamma_I$, further motivating the use of different $\gamma_I$ for
each instrument. All orbital parameter values are given for a given reference
epoch $t_0$, which we arbitrarily assume to be the first observation date.

Concerning the statistical description of the data, and under the same
assumptions as R14 (white noise, statistically independent measurements), we
define the logarithm of the likelihood function as
\begin{eqnarray} 
\ln L &=& - \frac{N_{\rm obs}}{2} \ln 2 \pi -
\frac{1}{2}\sum_{i=1}^{N_{\rm obs}} 
\ln \left(\epsilon_i^2+s_I^2\right)  -
\frac{1}{2}\sum_{i=1}^{N_{\rm obs}} 
\frac{\left(v_i - v\left[t,\vec{\theta}\right]\right)^2}{
\epsilon_i^2+s_I^2}\,, \label{eq:likelihood}
\end{eqnarray} 

\noindent which is our merit statistic to be maximized. The $s_I$ parameter is
often called \textit{jitter} and accounts for extra white-noise of each dataset
$I$. When $s_I=0$, maximizing this log-likelihood function is equivalent to
minimizing the $\chi^2$ statistic. When including correlation terms to the data,
the jitter parameter is even more necessary. That is, the index $\alpha$ also
has uncertainties and might include variability not traced by the Doppler
measurements. As for the constant offsets, $\gamma_I$, this extra-jitter will be
accounted for through $s_I$ of each dataset.

R14 proposed to split the Doppler series of GJ~581 in five chunks which
implicitly assume five independent sets (each one with a possibly different
$\gamma_I$, $C_I$ and jitter level $s_I$). Details and motivation for such
slicing are given in R14 and are based on apparent intervals of stronger
activity and the natural yearly sampling of the data. Consequences of applying
this slicing of the data are discussed later.

\subsection{Likelihood-ratio periodograms}\label{sec:periodograms}

A periodogram is a representation of the improvement of some merit statistic
when a sinusoidal signal is included in the model. Because of its numerical
efficiency, the most widely spread algorithm to compute periodograms is the
so-called Lomb-Scarge periodogram (or LS). The LS algorithm adjusts one sinusoid
to each test period resulting in a plot of the period (x-axis) against the
improvement of the $\chi^2$ statistic. A detailed derivation of the LS
periodogram from $\chi^2$ minimization is given in \citet{scargle:1982}. Under
the assumptions of the method, the peaks and significances derived from LS
periodograms are only representative of a test of significance when there are no
other signals present in the time-series. 

This problem can be circumvented by realizing that a periodogram can be used as
a representation of the overall model improvement \citep{baluev:2009,
anglada:2012c} by using a merit statistics of the complete model. That is, when
searching for evidence of a $k+1$ periodicity, we also need to simultaneously
adjust for all the other free parameters. This makes such \textit{periodograms}
computationally expensive, and one needs to create specific implementation of
the algorithms instead of using freely distributed tools. Periodogram procedures
based on adding one signal at a time are sometimes called \textit{hierarchical}
methods (detection is done from most significant to smaller signal). More
general methods that directly explore the full parameter space exist but will
not be discussed here for brevity. Some reported implementations of these
include tempered Markov Chain Markov Chain algorithms \citep{gregory:2011},
Delayed Rejection Adaptive Metropolis \citep[or DRAM][]{tuomi:corot7:2014} and
Markov Chains with nested sampling \citep{brewer:2015}.

\begin{figure} 
\includegraphics[angle=0, width=0.85\textwidth,clip]{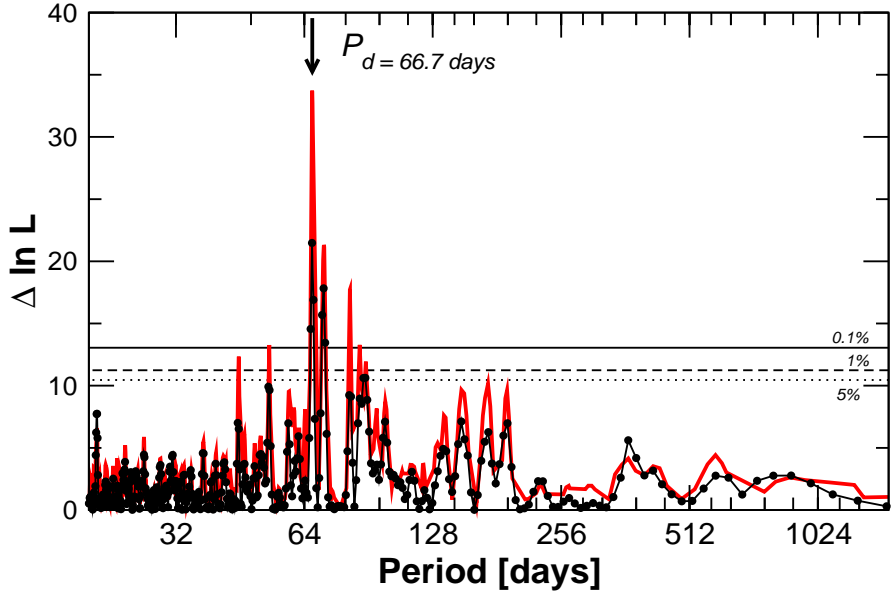} 
\caption{Likelihood ratio search for a 4th signal
demonstrating that the presence of GJ 581d is still strongly supported by the
data, despite the correlations with $H_\alpha$. }\label{fig:periodogram}
\end{figure}

In our analysis, we use our custom made software to perform optimization of the
likelihood function at the period search level. It differs from other Keplerian
fitting codes in the sense that allows adjusting correlation coefficients,
jitter parameters, and offsets as free parameters as well (further effects can
be easily incorporated when necessary). For example, note that the $C_I$
coefficients are linear parameters, so they can be trivially incorporated in a
general least-squares solver. All the parameters in $v[\vec{\theta}; t_i]$ are
converged using regular least-squares solving methods, and the parameters of the
likelihood (eg.~jitter terms s$_I$) are converged using steepest descent steps.
This process is iterated a few times until a small threshold $\delta \ln L$ is
registered between iterations. The solution is finally converged to the local
likelihood maxima (periodogram peak) using annealing. At the signal search
level, the $k+1$ signal is always considered sinusoidal (circular orbit) to
avoid problems with the non-linear behaviour of high eccentricities \citep[see
discussion in Appendix A in][]{anglada:2013}. Given that adding two more
parameters($e$ and $\omega$) can only improve the fit to the data, beating a
given significance threshold for the sinusoid provides a \emph{sufficient
condition for detection}. Significance assessments are finally provided using
False alarm probability (or FAP) estimates as described in \citet{baluev:2009,
baluev:2013}. FAPs smaller than 1\% usually imply significant detections and
more accurate significances can be later estimated using the integration of the
Bayesian posterior distribution \citep[eg.][]{tuomi:posterior:2012}. Our
complete $3+1$ planet model for five datasets contains 32 free parameters:
$3\times 5$ Keplerian ones, $3$ parameters for the $k+1$ sinusoid, and $3 \times
5$ parameters for the five subsets. The periods of the test sinusoids are
initialized over 8000 seed values uniformly sampled in frequency (1/P) between
1/1.1 and 1/20000 days$^{-1}$. The result of this procedure is illustrated in
Fig.~\ref{fig:periodogram}. Generating such periodogram took $\sim$ 30 min on a
standard 2.5GHz single-core CPU. More optimal implementations of likelihood
periodograms are given in \citet{planetpack}. 

\subsection{Results}\label{sec:results}

The Doppler time-series and I$_{H_\alpha}$ were used as provided in R14. Also as
in R14, one outlying I$_{H_\alpha}$ measurement was removed form the third
dataset (JD = 2454610.74293 days, likely caused by a flare). R14 only removes
correlations on three of the subsets based on apparently higher correlations. We
allow all five coefficients to be free parameters assuming that they can be
naturally zero if that value is preferred by the global fit.

As in R14, the three first signals at periods 5.3686, 12.914 and 3.1490 days
(GJ~581b, GJ~581c, and GJ~581e respectively) are easily detected despite the
correlations with I$_{H_\alpha}$. The likelihood-ratio periodogram search for
the 4-th signal (GJ~581d) is shown in Fig.~\ref{fig:periodogram} (red line). The
signal is well detected above the 0.1\% FAP line, implying a significant
detection beyond reasonable doubt. The black dots represent our periodogram
algorithm applied to the residuals to the 3-planet + correlations in an attempt
to replicate R14 analysis more closely. While this procedure shows lower
significance, we find that the significance of GJ~581d still does not fall below
the 1\% FAP threshold, suggesting additional relevant differences between our
model and R14 (eg., jitter is not optimized in R14, and it is unclear whether
constant offsets for each data chunk were solved as free parameters). In any
case, the much higher value of the red curve clearly shows that significance is
strongly boosted ($\Delta \ln L \sim 13$ implies $\sim e^{13}$ higher
significance) when all free parameters are adjusted. This feature is
characteristic of parameter degeneracy (activity signal is similar to the planet
candidate's one), but such partial degeneracy alone is not sufficient to negate
the significance of a much better model on statistical grounds. 
\begin{figure*}
\includegraphics[angle=0, width=0.95\textwidth, clip]{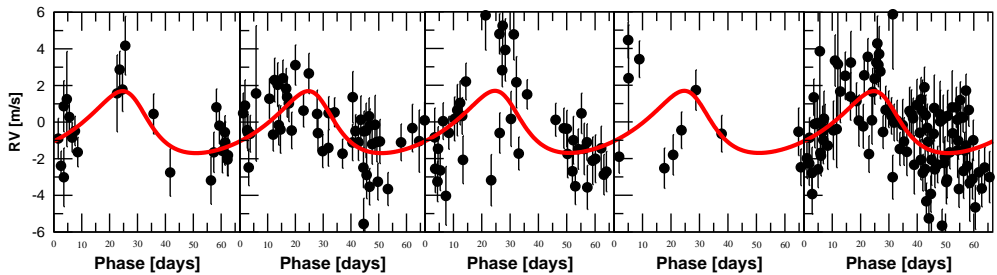}
\caption{Doppler measurements for each subset (chronological order from left to
right) after subtracting all parameterized effects (planets and correlations)
except GJ 581d and phase folded to a period of 66.7 days. The red line show the
maximum likelihood fit to the orbit. The signal is well traced in all the
subsets (2nd,3rd and 5th) with enough observations.}
\label{fig:phased}
\end{figure*}

\subsection{No evidence for correlations changing with
time}\label{sec:timevariability}

A time variable correlation can be easily explained by unrelated signals in both
RV and I$_{H_\alpha}$ in a similar period domain \citep[correlation does not
imply causality, see discussion in][]{velickovic:2015}. Just as an example,
Jupiter also has an orbital period (11.86 years) comparable to the activity
cycle of the Sun ($\sim$ 11 years). Unless the curves are in perfect phase, the
analysis of R14 would also detect time-dependent correlations. While skepticism
would be natural if only one cycle was covered, accumulated observations over
several cycles would clearly differentiate both signals (unless one keeps
adjusting a time-dependent correlation on arbitrary data-slices). The
Sun-Jupiter example can be compared to the period of 66.7 days of GJ~581d to the
signal in I$_{H_\alpha}$ at $\sim$ 130 days (and harmonics) discussed in R14.
A second consequence of forcing corrections into the Doppler time-series is the
increase in the noise floor for the RVs themselves. That is, Doppler time-series
become limited by the scatter in the activity indices, and -what is worse- they
can be severely contaminated by correlated variability of the indices as well. 

\begin{figure*}
\includegraphics[angle=0, width=0.95\textwidth, clip]{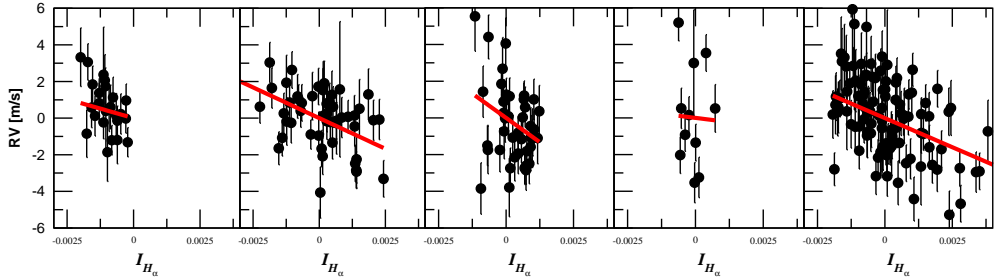}
\caption{Doppler versus H$_\alpha$ correlation plots for each subset. As for
the  phased plots in Figure \ref{fig:phased}, all signals except the
H$_alpha$/RV correlation have been subtracted to improve visualization. The
correlation slopes of the sets with more observations (2nd and 5th panel)
cast serious doubt on the proposed time-variability of the correlation law.
The values of the measured slopes are $C_1 = -420$ \ms, $C_2 = -670$ \ms,
$C_3 = -1040$ \ms, $C_4 = -190$ \ms and $C_5 = -630$ \ms, being $C_1$ and
and $C_4$ the smallest but also most uncertain given the small ammount of 
observations in those subsets.} 
\label{fig:correlations}
\end{figure*}

While we agree with R14 that correlations of RV/I$_{H_\alpha}$ are significant,
our analysis does not support the reported time-dependence of the
correlation either. Note that more discrepant slopes in Figure
\ref{fig:correlations} (Panels 1 and 4) correspond to seasons where the
number of points and the range of H$_\alpha$ variability is much smaller. Given
that the model is very complex (30+ parameters) and non-linear, proper
quantification of the uncertainties in the correlation coefficients requires
sampling techniques of the posterior density (eg. Monte Carlo Markov Chain
sampling of the posterior) which is beyond our scope here.

\subsection{Conclusions}\label{sec:conclusions}

The failure to confirm GJ~581d by R14 seems to be related to the analysis of
residual data and improper interpretation of periodograms.  R14 attempted
several correlations with activity indices and applied a rather arbitrary
slicing of the datasets. Furthermore, selecting apparently active sub-sets and
fitting correlations to those should be avoided as it constitutes a circular
argument. The same problem likely explains the non-detection of the very
significant signal GJ~667Cd in \citet{robertson:2014b}, even under the
assumption of white noise. In that case, the authors sliced the RV time-series
and forced correlations with another activity index (the so-called FWHM). As for
GJ~581d -and Jupiter for the Sun-, GJ~667C also show evidence for variability in
periods comparable ($P_{\rm FWHM}\sim 105$~days) to the period of the proposed
planet candidate ($P_{d}=91$~days). Again, removing time-dependent correlations
on data slices necessarily decrease their apparent significance, especially in
periodograms of residual data.

The validity of the various models and methods to account for noise in Doppler
data is a hotly debated topic. Contributions to the discussion on benchmark
systems should ensure that the applied statistical tools are formally correct.
Given all these caveats, the analysis presented in R14 should be considered
inconclusive, and GJ~581d should be reinstated as a planet candidate until
additional observations suggest otherwise.

\bibliography{biblio}
\label{lastpage}

\end{document}